\documentclass[%
superscriptaddress,
amsmath,amssymb
pra,twocolumn,
floatfix
]{revtex4-1}

\usepackage{graphicx}
\usepackage{dcolumn}
\usepackage{bm}
\usepackage[usenames]{color}

\usepackage{xcolor}
\usepackage{amssymb,amsmath}
\usepackage[normalem]{ulem}

\begin{document}
\title{Nonlinear magneto-optical rotation with parametric resonance}
\date{\today}

\author{P. Put}
\affiliation{M. Smoluchowski Institute of Physics, Jagiellonian University, {\L}ojasiewicz 11, 30-348 Krak\'ow, Poland}

\author{P. Wcis\l o}
\affiliation{Institute of Physics, Faculty of Physics, Astronomy and Informatics, Nicolaus Copernicus University, Grudziadzka 5/7, 87-100 Toru\'{n}, Poland}

\author{W. Gawlik}
\affiliation{M. Smoluchowski Institute of Physics, Jagiellonian University, {\L}ojasiewicz 11, 30-348 Krak\'ow, Poland}

\author{S. Pustelny}
\email{pustelny@uj.edu.pl}
\affiliation{M. Smoluchowski Institute of Physics, Jagiellonian University, {\L}ojasiewicz 11, 30-348 Krak\'ow, Poland}

\begin{abstract}
We report on investigations of nonlinear magneto-optical rotation (NMOR) in rubidium vapor subjected to a modulated magnetic field and continuous-wave (CW) laser-light illumination. By superimposing modulation and a static (DC) magnetic field, we demonstrate the appearance of resonances at both small and large (compared to the ground-state relaxation rate) values of the static field. Since in conventional NMOR, there is no rotation at high fields, this suggests an existence of a novel mechanism generating anisotropy in the considered case, which we identify as parametric resonance. The experiments are performed using light of small ellipticity and rotation signals are significantly enhanced by combining atom-induced polarization rotation with a passive rotation induced with a wave plate.
All the observations are supported with theoretical simulations. The density-matrix formalism and angular-momentum probability surfaces are used to provide intuitive explanation of the observed signals.
\end{abstract}
\maketitle

\section{Introduction}
Parametric resonance is a phenomenon known across many disciplines, including physics, electronics, mechanics, and biology. The effect occurs when one of system parameters, e.g., resonance frequency or decay rate, is modulated \cite{Landau}. In the context of ligh-matter interactions, the parametric resonance was first investigated in the experiments on optical pumping and radio-frequency (rf) spectroscopy by Alexandrov \textit{et al.} \cite{Aleksandrov63}. The authors found that modulation of a static magnetic field with significantly weaker rf field resulted in appearance of narrow resonances in the intensity of transmitted light. The resonances arose when the rf-field frequency matched the Larmor frequency of illuminated atoms subjected to the static field. Cohen-Tannoudji and coworkers interpreted the phenomenon in terms of a dressed-atom formalism \cite{Cohen70}. These studies eventually led to the development of a sensitive magnetometric technique \cite{Dupont70}. Novikov and Malyshev performed another parametric-resonance experiment, where a relaxation rate of $^{133}$Cs atoms was varied by application of  modulated, resonant, and unpolarized light \cite{Novikov1975}. Modulation of the atoms' relaxation rate resulted in appearance of parametric resonances observed in transmission as a function of the magnetic field. These results were in good agreement with a theoretical treatment developed by Okunevich \cite{Okunevich1974}.
The advent of lasers brought new dimension into the studies of parametric resonances. Particularly, Failache \textit{et al.} \cite{Failache03} applied the parametric resonance for verification of the theory of Taichenachev \textit{et al.} on electromagnetic induced absorption (EIA) \cite{Taichenachev00}. According to the theory, in a degenerate two-level system, EIA arises due to coherence transfer from the excited state to the ground state. By measuring intensity of light transmitted through an atomic vapor, the authors demonstrated a sign reversal of the observed transmission resonance (enhanced transmission versus enhanced absorption) and thus confirmed the theory. The parametric resonance was also investigated in the context of magnetometry within the so-called spin-exchange-relaxation-free regime \cite{Selzer2004Unshielded,Li2006Parametric}, where modulation of the magnetic field transversely to the light-propagation direction allowed determination of the transverse-field components. Moreover, parametric resonance became a useful method for characterisation of cold-atom/molecule traps by the means of so-called trap-loss spectroscopy \cite{stwalley1999photoassociation, krems2009cold}.

Here, we present the studies of nonlinear magneto-optical rotation (NMOR) \cite{Budker2002RMP}, i.e., the effect of light polarization rotation in a medium subjected to a magnetic field \footnote{Note a broader definition of the effect also including elliptically polarized light, while typically the phenomenon only concerns rotation of polarized light.}. In conventional NMOR experiments, when CW light is used, nonzero polarization rotation is observed only at weak fields $|\omega_L^{(0)}|\lesssim\Gamma_g$, where $\omega_L^{(0)}$ is the Larmor frequency due to the DC field and $\Gamma_g$ is the ground-state relaxation rate \cite{Budker1998}. While this opens applications of the effect to sensitive magnetometry, it also limits the technique for the detection of weak fields only. This constraint may be alleviated by implementing modulation of the light frequency \cite{Budker2002}, amplitude \cite{Gawlik2006}, or polarization \cite{Weis2013}, which extends the dynamic range of NMOR magnetometers by several orders of magnitude (up to the level of the Earth magnetic field \cite{Acosta2006}), offering ultra-precise measurements in a broad dynamic range \cite{Pustelny2008}. In this work, we demonstrate that a similar effect may be obtained with CW light by modulating longitudinal magnetic field. We refer to this effect as Parametric NMOR (PNMOR). The modulation of the field leads to the appearance of resonances, whose positions are determined by the static magnetic field. We study the effect under different physical conditions and based on theoretical analysis using the density-matrix (DM) formalism, we demonstrate a strong competition between optical pumping and light-induced transverse relaxation processes. Visualizing the DM with angular-momentum probability surfaces \cite{Rochester2001Atomic} facilitates an intuitive interpretation of the effect and provides an insight into the PNMOR effect, which enables  future applications to sensitive high-field magnetometry.

The article is organized as follows. In the next section, we describe the experimental setup. Then, we introduce our theoretical approach (Sec.~\ref{sec:TheoryBrief}). The results of theoretical modeling along with experimental data are presented in Sec.~\ref{sec:Results}. We start with a brief review of conventional NMOR (Sec.~\ref{sec:NMORtypicalFirst}) and then present our main results. First, the discussion of PNMOR with alternating (AC) field significantly exceeding DC magnetic field is presented (Sec.~\ref{sec:NMORweak}). Next, we discuss situation when comparable AC and DC magnetic-field components, both largely exceeding the ground-state relaxation rate, are used (Sec.~\ref{sec:NMORstrong}). The final remarks and conclusions are gathered in Sec.~\ref{sec:Conlusions}. 

\section{Experimental setup\label{sec:Setup}}
\begin{figure}
\includegraphics[width=\columnwidth]{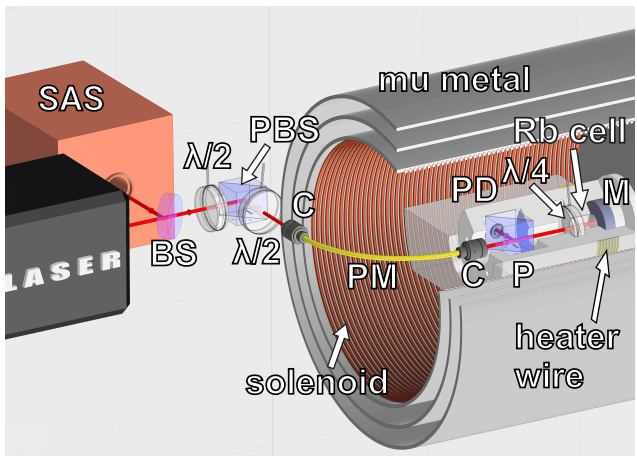}
\caption{Layout of the experimental setup used for measurement of PNMOR. Resonant laser beam is marked by bright red lines in the diagram. Wavelength monitoring is done with the saturated absorption spectroscopy (SAS) system (orange box in the diagram). BS denotes the beam splitter, PBS marks the polarizing beam, PM is the polarization maintaining fiber. A half-wave plate ($\lambda/2$) situated before PBS is used to control the beam intensity, while the second half-wave plate provides appropriate polarization orientation along the fiber axis. Fiber couplers are denoted by C, while polarizer is marked with P. Light directed to a sensor head (white cylinder in the diagram) is sent through a quarter-wave plate ($\lambda/4$), rotated by a small angle $\alpha$ ($3^\circ-7^\circ$) with respect to the initial light polarization, acquiring small ellipticity. Next, the light illuminates the paraffin-coated vapor cell (Rb cell) and passes the cell for the second time being retroreflected on the mirror (M). Finally, a fraction of the light is directed to the photodiode (PD) where it is detected.}
\label{fig:Setup}
\end{figure}
A paraffin-coated cylindrical glass cell of a volume of $\approx1.1$~cm$^3$ (length of 1.3~cm, diameter of 1.1~cm), containing isotopically enriched sample of $^{87}$Rb was placed inside a three-layer semi-open $\mu$-metal magnetic shield, reducing stray fields by a factor larger than 10$^3$. The cell was electrically heated up to 50$^\circ$C (corresponding to a concentration of about $10^{11}$~atoms/cm$^3$) using a bifilar nonmagnetic wire (the heater generated negligible magnetic field inside the cell). A solenoid of an effective length (increased by magnetostatic imaging in the innermost $\mu$-metal layer) of 1.6~m was used to generate homogenous magnetic field along the $z$ axis. The field was controlled using a DC current source and a function generator.

Rubidium vapor was illuminated with light emitted by a homemade extended-cavity diode laser. The light was tuned to the $F=2\rightarrow F'=1$ transition of the rubidium D1 line (795~nm) and its frequency was monitored with saturated-absorption spectroscopy. Light intensity was controlled in a range between 1--1000 $\mu$W by a half-wave plate and a polarizing beam-splitter, situated in front of a polarization-maintaining fiber. The fiber was used to spatially filter the light beam and bring it to the vapor cell. Before the cell, the light beam (2-mm in diameter) was polarized with a high-quality crystal polarizer, and then sent through a quarter-wave plate, with an optical axis rotated by the angle $\alpha$ with respect to the polarizer optical axis. After the cell, the beam was retroreflected, so that it traversed the cell and waveplate twice. Double pass through the rotated quarter-wave plate resulted in an accumulated phase shift of $\pi$ and passive rotation of the light polarization by 2$\alpha$. Polarization rotation of the beam was detected by monitoring the intensity of light directed in a side channel of the crystal polarizer. For small rotation, the intensity $I$ of detected light is given by:

\begin{equation}
I=I_0\sin^2\left[2\alpha+A(t)\right]\approx I_0\left[4\alpha^2+4\alpha A(t)+A^2(t)\right],
\label{eq:TransmittedIntensity}
\end{equation}
where $I_0$ is the intensity of incident light and $A(t)$ is the time-dependent angle of polarization rotation induced by atoms (the NMOR signal). The photodiode output current was detected using a lock-in amplifier, operating at the first harmonic of the magnetic-field modulation frequency $\omega_m$ (except for the experiment described in Sec.~\ref{sec:NMORtypicalFirst}). 
Here, we wish to stress a crucial role of the quarter-wave plate and double-pass configuration. As in most NMOR experiments polarization rotation rarely exceeds 10~mrad \cite{Budker2002,Pustelny2008,pustelny2006pump,ledbetter2007detection}, the homodyne detection employed here results in a significant enhancement of the rotation-detection capabilities. Moreover, the dependence of $A(t)$ on a different harmonic of the modulation frequency allows one to eliminate other terms of Eq.~\eqref{eq:TransmittedIntensity} by demodulating the detected signal at a specific harmonic of the modulation frequency $\omega_m$ and reduce spectral noise of the signal (resulting in reduced contribution from 1/$f$ noise).

\section{Theoretical background\label{sec:TheoryBrief}}

Despite complex energy structures of real atomic systems used in NMOR, many distinct features of the effect may be modelled using a simple two-level system with the total angular momentum $F=1$ of the ground state and $F'=0$ of the excited state \footnote{In the text the subscripts of the density matrix elements refer to Zeeman sublevels of the ground state (marked with -1,0,1) and to one excited level if marked with 0'.}. We make use of such a simplified $\Lambda$-system to investigate the appearance of PNMOR with modulated magnetic field associated with the Larmor frequency given by
\begin{equation}
\omega_L(t)=\omega^{(0)}_L+\Delta\omega_L\cos\omega_m t,
\label{eq:MagneticField}
\end{equation}
where $\omega_{L}^{(0)}$ corresponds to the DC magnetic-field component, while $\Delta\omega_L$ and $\omega_m$ are the amplitude and frequency of the modulated (AC) part of the field respectively. 

The NMOR angle $\varphi$ may be calculated using the DM formalism \cite{Happer2010}. The detailed derivation of the relevant equations was presented in Ref.~\cite{Pustelny2011Tayloring}. To avoid repetitions, here we skip the details and summarize the main points of the theoretical analysis preserving the notation of the Ref.~\cite{Pustelny2011Tayloring}. In systems with a quantization axis oriented along the light-propagation direction ($\sigma$-polarized light), the NMOR angle $\varphi$ can be expressed as
\begin{equation}
\varphi\propto\frac{1}{\varepsilon}\text{Re}(\widetilde{\rho}_{-1,0'}-\widetilde{\rho}_{1,0'}),
\label{eq:RotationDMFirst}
\end{equation}
where $\varepsilon$ is the amplitude of electric field of light and $\widetilde{\rho}_{-1,0'}$ and $\widetilde{\rho}_{1,0'}$ are the slowly-varying amplitudes of optical coherences induced by the $\sigma^{+}$- and $\sigma^{-}$-components of $\sigma$-polarized light, respectively. 

Conceptually, NMOR can be considered as a three-stage process.  Initially, atoms are subjected to (pump) light, which redistributes population between energy levels and creates coherences between them. Next, an atomic state evolves due to other (e.g., magnetic) fields.  Finally, the atom state is probed by light which results in polarization rotation. Physically, both pumping and probing can be accomplished by the single beam. In contrast to the first interaction however, the probe interaction is linear and hence, to the first approximation, it does not affect the state of the atoms.

From a theoretical standpoint, the first two processes can be described simultaneously.  This is done by incorporating both optical- and magnetic-field interactions into the interaction Hamiltonian $H_i$
\begin{equation}
    H_i=H_l+H_B=-\vec{E}\cdot\vec{d}-\vec{\mu}\cdot\vec{B},
    \label{eq:Hamiltonian}
\end{equation}
where $H_l$ and $H_B$ are light- and magnetic-interaction Hamiltonians, respectively, $\vec{d}$ and $\vec{\mu}$ are electric and magnetic dipole moments, respectively, $\vec{B}$ is the magnetic field and $\vec{E}$ denotes the electric-field of light.  By substituting the Hamiltonian into the Liouville equation
\begin{equation}
    \dot{\rho}=-\dfrac{i}{\hbar}\left[H_0+H_i,\rho\right]+\Gamma(\rho)+\Lambda,
    \label{eq:LiouvilleEquation}
\end{equation}
where $H_0$ is the unpertubed Hamiltonian, $\Gamma(\rho)$ is the relaxation operator including all relaxation processes, $\Lambda$ is the DM in thermal equilibrium.
When Hamiltonian \eqref{eq:Hamiltonian} depends on time only through a harmonic dependence of the electric-field of light, the steady-state DM can be found by zeroing all left-handed sides of the Liouville equations (after application of the rotating-wave approximation). In the case of a time-dependent Hamiltonian the evolution equation can be solved by a decomposition of the DM into a Fourier series versus the modulation frequency $\omega_m$ (see, for example, Ref.~\cite{Pustelny2011Tayloring}). In turn, one obtains an infinite set of the self-coupled differential equations which, if the series converges, may be truncated at a finite Fourier-expansion coefficient.  Specific coefficients correspond to DM elements oscillating at given harmonics of the modulation frequency $\omega_m$. Assuming that the quasi-steady state is reached \footnote{In this paper, a quasi-steady state is used to describe a situation when system parameters may vary over time but character of the evolution is periodic with stable frequency composition (stable amplitude and frequency).} by the system (meaning one can zero all time derivatives), the problem reduces to solving a set of algebraic equations. Although this approach enables derivation of the explicit formulae for the DM elements at specific harmonics \cite{Pustelny2011Tayloring}, it does not allow for the transient dynamics of the system to be investigated. For such analysis a numerical approach appears more useful. 

While optical pumping is a nonlinear interaction, probing anisotropy of the medium, which manifests as polarization rotation, is a linear process. Therefore, it is convenient to consider this process using a perturbative approach versus the probe-light amplitude. In the first order of perturbation, the optical-coherence amplitude $\widetilde{\rho}_{\mp 1,0'}^{(1)}$ is given by
\begin{equation}
\widetilde{\rho}_{\mp 1,0'}^{(1)}=\frac{\Omega \left(\rho_{\pm 1, \pm 1}^{(0)}+\rho_{\mp 1, \pm 1}^{(0)}\right)}{\Delta\omega_p+i\Gamma_e},
\label{eq:AmplitudeCoherencePerturbation}
\end{equation}
where $\Delta\omega_p$ is the detuning of the probe light from the atomic transition, $\Gamma_e$ is the excited-state relaxation rate and $\Omega$ is the probe-light Rabi frequency. Since interaction with a weak probe light occurs with the rate orders of magnitude lower than the spontaneous emission rate, we can safely assume that the equilibrium excited-state population $\rho_{0',0'}$ is zero ($\rho_{0',0'}^{(0)}=0$). At the same time the ground-state polarization (effect of population imbalance $\rho_{\pm 1, \pm 1}^{(0)}$ and coherences $\rho_{\mp 1, \pm 1}^{(0)}$) builds up until the interaction and contributes to the first-order optical coherence.

Substituting Eq.~\eqref{eq:AmplitudeCoherencePerturbation} into Eq.~\eqref{eq:RotationDMFirst} allows us to write the first-order optical rotation experienced by the probe beam
\begin{equation}
    \varphi^{(1)}\propto\frac{1}{\varepsilon_p}\textrm{Re}\left(\rho^{(0)}_{+1,+1}-\rho^{(0)}_{-1,-1}+\rho^{(0)}_{-1,+1}-\rho^{(0)}_{+1,-1}\right),
    \label{eq:PolarizationRotationTimeDependent}
\end{equation}
where $\varepsilon_p$ is the probe-light amplitude.  The dependence of the polarization rotation only on the ground-state population and coherences implies that it suffices to focus on the evolution of the ground-state part of DM to reliably describe the evolution of the system (in particular, Figs.~\ref{fig:ParametricLow} and~\ref{fig:ParametricHigh1} visualise only the ground-state). 

We present a temporal evolution of the DM in a form of three-dimensional angular-momentum probability surfaces \cite{Rochester2001Atomic}. The 3D structures are generated by projecting angular momentum in all spatial directions. This approach reveals spatial symmetry of the DM and the optical anisotropy of a medium, and is used to calculate the NMOR signals [Eq.~\eqref{eq:RotationDMFirst}].

Finally, in order to calculate polarization rotation at a given harmonic of the modulation frequency, the time-dependent polarization rotation $\varphi(t)$ [Eq.~\eqref{eq:PolarizationRotationTimeDependent}] needs to be multiplied by the sine or cosine function of the modulation-frequency multiplicity and then numerically integrated over the oscillation period (pseudo-lock-in detection). 

\section{Results and Discussion\label{sec:Results}}
\subsection{NMOR with DC magnetic field\label{sec:NMORtypicalFirst}}
\begin{figure}
\includegraphics[width=\columnwidth]{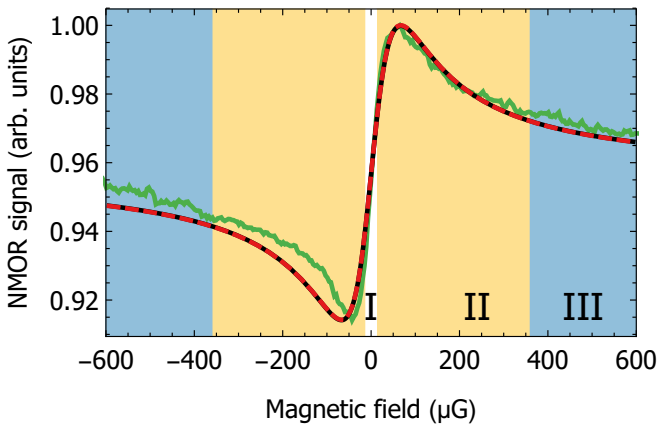}
\caption{(Color online) NMOR signals with elliptically-polarized light measured (green) and simulated (red and black) as a function of the unmodulated magnetic field $\omega_L^{(0)}$ ($\Delta\omega_L=0$). The three colors, white (I), light yellow (II), and blue (III), indicate regions of NMOR discussed in the text. Experimental data were measured with the light intensity of $\approx$~220~$\mu$W and the quarter-wave plate rotated by $\alpha = 3^\circ$. The simulations were performed for $\Gamma_e/\Gamma_g=10^{4}$, $\Omega/\Gamma_g=2.5\times 10^{2}$ and $\Delta\omega_L=0$. To reproduce experimental conditions, the simulations were performed with light of small ellipticity (dotted black line). For comparison, an additional signal with no ellipticity was simulated (dashed red). This signal is almost identical as in the previous case showing that the effect of ellipticity is negligible. A vertical offset of the measured and simulated data arises from the passive rotation $2\alpha$ caused by the double light pass through the quarter-wave plate.}
\label{fig:NMORtypical}
\end{figure}

We start the discussions by recalling standard NMOR. Figure~\ref{fig:NMORtypical} shows a typical NMOR signal measured as a function of unmodulated (pure DC) magnetic field ($\Delta\omega_L=0$) with CW light. This case will serve as a reference for discussion of the PNMOR signals. 

In the measurement shown in Fig. \ref{fig:NMORtypical}, one can clearly distinguish three different regimes of NMOR: (I) no rotation ($\varphi=0$) at $B=0$ [there is only a static offset corresponding to the background term $\alpha^2$ in Eq.~\eqref{eq:TransmittedIntensity}], (II) strong dependence of rotation on magnetic field for $|\omega_L^{(0)}|\lesssim 3 \Gamma_g$, and (III) no rotation for the strong field ($|\omega_L^{(0)}|\gg\Gamma_g$). 

Rotation of the major axis of the elliptical light polarization, observed at fields close to, but not exactly zero (region II), arises from pumping of atoms with nearly linearly polarized light and their subsequent Larmor precession. Motion of atoms causes the process of optical pumping to occur at random times, resulting in atoms with polarizations representing many spatial orientations. However, the number of atoms with a given orientation depends on the rotation angle, as larger rotations requires longer evolution time during which atoms relax toward thermal equilibrium. Consequently, the net anisotropy of the medium depends on the magnetic field and so does the rotation angle. At zero field (region I), when medium's polarization is the strongest, there is no Larmor precession, so that the medium anisotropy axis is parallel to the major semi-axis of light and no rotation is observed. For the strongest fields (region III), the Larmor precession is so fast that the atomic polarization rotates many times before relaxing, which results in washing out of the transverse polarization of atoms (longitudinal polarization is preserved) hence destroying the transverse optical anisotropy. Before reaching the steady state described above, the system undergoes transient evolution with distinct NMOR signals \cite{grewal2018transient}.
\subsection{PNMOR with weak DC fields\label{sec:NMORweak}}

\begin{figure}
\includegraphics[width=\columnwidth]{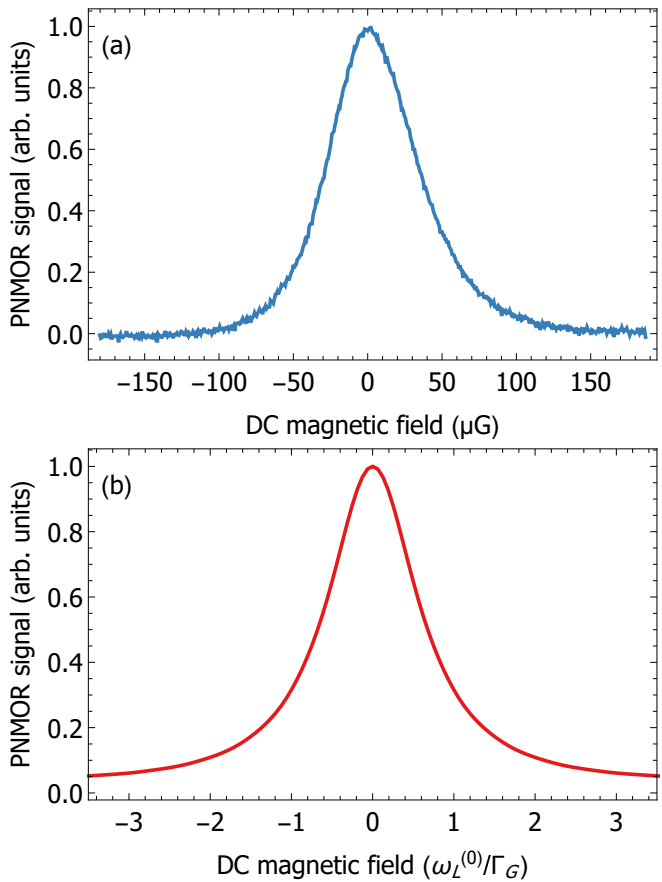}
\caption{(a) Experimentally measured and (b) numerically simulated amplitude of the PNMOR signal as a function of the DC magnetic field $\omega_L^{(0)}$ with the strong AC component ($\Delta\omega_L\gg\Gamma_g$). The experiment was performed for modulation amplitude $\Delta \omega_L/2\pi=5\times 10^3$~Hz ($7.1$~mG), modulation frequency of $2 \times 10^4$~Hz, and laser power of $\sim40$~$\mu$W. For numerical calculations the following values of the parameters were used: $\Gamma_e/\Gamma_g=10^{4}$, $\Omega/\Gamma_g=1.5 \times 10^{2}$, $\omega_m/\Gamma_g=4 \times  10^{2}$ and $\Delta\omega_L/\Gamma_g=10^{2}$.}
    \label{fig:ParametricLowExp}
\end{figure}

Next, we investigate PNMOR with weak DC ($|\omega_L^{(0)}|\lesssim5\Gamma_g$) field component. Figure~\ref{fig:ParametricLowExp}(a) presents the amplitude of the PNMOR signal (phase-sensitive detection, $R$ signal) measured versus the DC field $\omega_L^{(0)}$. The observed dependence is characterized by an absorptive Lorentzian curve centered at $\omega_{L}^{(0)}=0$, indicating that maximum rotation is observed at zero field. For the experimental parameters, the resonance has a width (full width at half maximum) of 50~$\mu$G, which is limited by the quality of antirelaxation coating, atom-atom collisions (spin-exchange and spin-randomization) and power broadening. Results of corresponding DM calculations are shown in Fig.~\ref{fig:ParametricLowExp}(b). In this case, a dynamic PNMOR signal in the quasi-steady state is numerically integrated, and the "lock-in signal" is obtained. The calculated signal has the same shape (absorptive peak centered at $\omega_L^{(0)}$) and width as its experimental counterpart, validating the correctness of used experimental model.

An important difference between NMOR and PNMOR is nonzero rotation observed for larger values of instantaneous magnetic fields. While in conventional NMOR, when the Larmor frequency $\omega_L(t)$ greatly exceeds the ground-state relaxation rate $\Gamma_g$ there is no rotation (region III), in PNMOR, strong signals with amplitudes corresponding to those of the zero-field NMOR are observed. This difference clearly indicates that a distinct mechanism is responsible for generation of NMOR in the two cases.

In order to theoretically investigate the effect, we analyze evolution of the DM due to the oscillating longitudinal magnetic field. For simplicity, we assume zero DC field component, $\omega_L^{(0)}=0$, which corresponds to maximum polarization rotation (Fig.~\ref{fig:ParametricLowExp}), but analogous analysis can be also performed at non-zero DC fields, $\omega_L^{(0)}\neq0$. The simulations are performed under the quasi-steady state conditions, i.e., for $t\gg1/\Gamma_g,1/\Gamma_e,1/\Omega,1/\Delta\Omega_L,1/\omega_m$, within a single magnetic-field modulation period $T_m=2\pi/\omega_m$. Analysis of the evolution of the DM in such a period fully captures the system dynamics.

\begin{figure}
\includegraphics[width=\columnwidth]{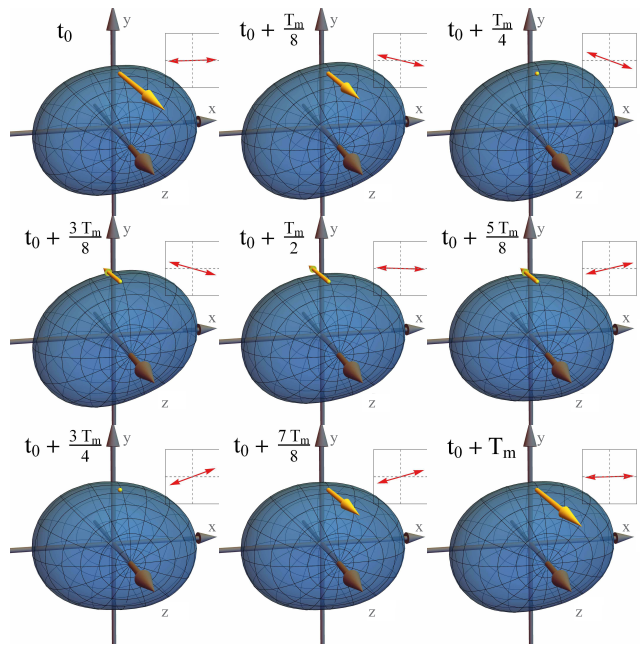}
\caption{(Color online) Evolution of the ground-state DM under interaction with light of slight ellipticity, zero DC field and oscillating longitudinal magnetic field. The yellow arrow represents instantaneous magnetic field. The inset shows calculated polarization rotation (angle of rotation has been increase 50-fold for better visibility). The simulations were performed for $\omega_m=1/4~\Delta\omega_L$, $T_m=2\pi/\omega_m$, $\Omega=1.5\cdot 10^{2} \Gamma_g$, $\Delta\omega_L=10^{2}\Gamma_g$, and $\omega_L^{(0)}=0$.}
    \label{fig:ParametricLow}
\end{figure}

As outlined in Sec.~\ref{sec:TheoryBrief}, light of a small ellipticity pumps atoms into a state corresponding to axial orientation of atomic angular momentum (atomic alignment) with a small asymmetry between two directions (small admixture of orientation). With the DM visualization technique, the alignment manifests as an ellipsoid-like shape elongated along the ellipticity axis (Fig.~\ref{fig:ParametricLow}).
In contrast to standard NMOR, however, even at $\omega_L^{(0)}=0$ the DM is not static but evolves due to the oscillating part of the magnetic field $\Delta\omega_{L}$. This temporal evolution makes PNMOR more similar to NMOR with modulated light at magnetic fields stronger than conventional NMOR with CW light and low magnetic field  \cite{Budker2002,Gawlik2006,Weis2013}. Nonetheless, even in this case, the system evolution becomes eventually periodic with repeating pumping and precession periods. In particular, for the DM corresponding to the ellipsoid oriented along the light polarization (at times $t=t_0+m/2~T_m$, where $m$ is an integer), optical pumping constructively contributes to the overall polarization (former and newly generated atomic polarizations are parallel). For nonparallel orientations of light polarization and atomic alignment, light tends to repump atoms into a different state (polarization along light polarization), destroying atomic transverse polarization. This repumping process introduces extra relaxation and reduces rotation at higher light power \footnote{This is the same mechanism that, in conventional NMOR, bring polarization rotation to zero at higher magnetic fields (Fig.~\ref{fig:NMORtypical}, regime III).}. Since the light-induced relaxation varies over time due to modulation of the magnetic field, the optical anisotropy is generated in the system and the polarization rotation is observed due to parametric resonance.

\begin{figure}
\includegraphics[ width=\columnwidth]{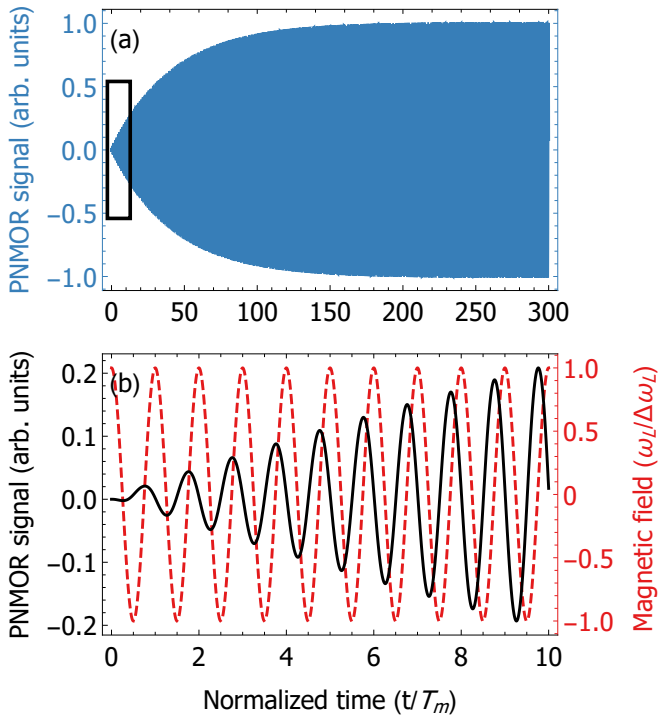}
\caption{(Color online) Simulated dependence of PNMOR signal on longer (a) and shorter (b) time scales. The black rectangle corresponds to evolution over ${10~T_{m}}$. The solid black line is the $\varphi(t)$ signal and the red dashed line is the trace of the instantaneous magnetic field The simulations were performed for $\omega_L^{(0)}=0$ and other parameters the same as in Fig.~\ref{fig:ParametricLowExp}.}
\label{fig:OpticalCoherencesLow}
\end{figure}

To further investigate the process of anisotropy generation, we investigated the system dynamics starting from thermal equilibrium (equal populations of the ground-state sublevels and no coherences) after turning on the light and magnetic field modulation. Figure~\ref{fig:OpticalCoherencesLow} shows the time evolution of the PNMOR signal under the same conditions as before.
The amplitude of these oscillations increases over time, eventually stabilizing at the value determined by the light intensity, ground-state relaxation rate, and strength of the DC magnetic field [Fig.~\ref{fig:OpticalCoherencesLow}(a)]. Magnification of the initial part of the oscillations [Fig. \ref{fig:OpticalCoherencesLow}(b)] shows that polarization starts to oscillate right after $t=0$ at the modulation frequency $\omega_m$. The oscillations are also phase shifted with respect to the magnetic-field modulation.

\begin{figure}
\includegraphics[width=\columnwidth]{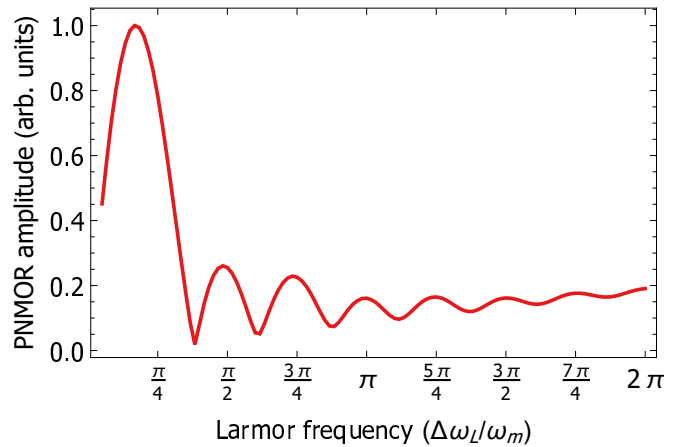}
\caption{Simulated dependence of the low-field PNMOR signal amplitude on the modulation frequency. The simulation was performed for the same parameters as those used in Fig.~\ref{fig:ParametricLowExp}.}
    \label{fig:NMORfreq}
\end{figure}

An important question about PNMOR concerns the dependence of the amplitude of the observed signal on the modulation frequency. To investigate this in more detail, we simulated the dependence of the PNMOR amplitude on the ratio $\Delta\omega_{L}/\omega_m$(Fig.~\ref{fig:NMORfreq}). The dependence reveals that for low modulation frequency, the signal increases with the frequency, reaching its maximum at $\Delta\omega_L/\omega_m=\pi/4$. For further increase of $\omega_m$, the amplitude decreases with $\omega_m$ (up to about $\Delta\omega_L/\omega_m=3\pi/8$, when another maximum starts to build up). To understand this behavior, we recall that the strongest polarization rotation is observed when the anisotropy axis and light polarization are oriented by 45$^\circ$. Even for the same degree of the medium polarization, the smaller angles correspond to weaker NMOR signals ($\varphi=0$ for parallel orientations). Similarly, for larger rotation angles the NMOR signal is also smaller. From this it follows that the phase accumulated between two extreme positions (two turning points) has to be equal to $\pi/2$. The phase accumulated between two extreme anisotropy orientations at times $t_1$ and $t_2$, is given by
\begin{equation}
    \begin{split}
	\Delta\phi(t_2,t_1)=&\int_{t_1}^{t_2}\Delta\omega_L\cos(\omega_m t)dt=\\
    =&
	\frac{\Delta\omega_L}{\omega_m}\left[\sin(\omega_{m}t_2)-\sin(\omega_{m}t_1)\right],
	\end{split}
\end{equation}
where for simplicity we assumed $\omega_L^{(0)}=0$. Thus, assuming that in the simplest case the turning points occur at $t_1=\pi/(2\omega_m)$ and $t_2=3\pi/(2\omega_m)$, one can show that the most efficient pumping and hence the strongest rotation occurs at $\Delta\omega_L/\omega_m=\pi/4$. Even though the amplitude reaches maximum under such conditions, the frequency dependence of PNMOR is relatively weak. Specifically, a change in the modulation frequency of $10\%$ corresponds to a roughly $5\%$  variation of the amplitude.  In Fig.~\ref{fig:NMORfreq} one can also observe additional maxima at higher amplitudes of modulation, with ratio $\Delta\omega_L/\omega_m$ equal to multiples of $\pi/4$, albeit with decreasing amplitude. These maxima correspond to accumulated phases $\varphi=\pi/2+m\pi$, where the $\pi$-factor reflects two-fold symmetry of the medium anisotropy.

\begin{figure}
\includegraphics[width=\columnwidth]{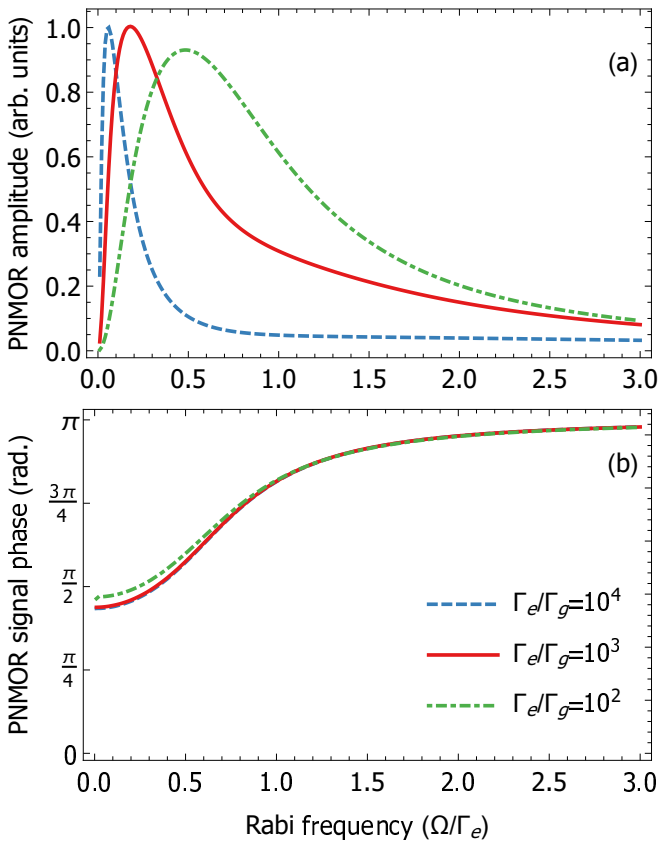}
\caption{(Color online) (a) Amplitude and (b) phase of the PNMOR simulated for $\omega_L^{(0)}/\Gamma_g=0$ as a function of the normalized Rabi frequency $\Omega/\Gamma_e$. These calculations were carried out for $\omega_m/\Gamma_e= 10^{-1}$ and $\Delta \omega_L/\Gamma_e=2.5\times 10^{-2}$.}
    \label{fig:ParametricLowPhaseAmpl}
\end{figure}

To gain more insight into the PNMOR mechanism, properties of the resonance are studies versus light intensity. Figure~\ref{fig:ParametricLowPhaseAmpl}(a) presents the signal amplitude dependence on light intensity in the quasi-steady state, for three distinct relaxation-rate ratios.
In the simplest case, i.e., when both the Rabi frequency $\Omega$ and the ground-state relaxation rate $\Gamma_g$ are smaller than the amplitude of the magnetic-field modulation $\Delta\omega_L$, the dynamics of the atomic polarization and polarization rotation is predominantly determined by the magnetic interaction (the effect of dephasing due to optical repumping is negligible). Therefore, an increase of light intensity corresponds to more efficient generation of medium anisotropy. However, at higher light intensities, the repumping of atoms becomes more prominent and reduces overall anisotropy of the medium and the rotation amplitude. This behavior is also present when a phase delay between modulation and polarization rotation is analyzed [Fig.~\ref{fig:ParametricLowPhaseAmpl}(b)]. For weak light, the phase shift equals nearly $\pi$/2 (small difference between three plots originates from the competition between the effect of optical pumping and relaxation of atomic-polarization). This shift can be explained by noting that positive precession is observed for positive fields, even after passing the maximum value of the instantaneous field $\omega_L(t)$, but before changing the field direction, the angle of polarization rotation keeps increasing and reaches its maximum at the moment of magnetic-field direction reversal. In turn, there is $\pi/2$ phase shift between  the magnetic-field modulation and polarization rotation. For more intense light ($\Omega>\Delta \omega_L$), the optical repumping starts to play an important role, and decreasing instantaneous magnetic field not only slows down the Larmor precession but also enhances reorientation of the net anisotropy due to increased contributions from atoms oriented along the light polarization \footnote{This is true for anisotropy axis rotated by less than $\pi$/4 at any time. Such conditions are true for most of NMOR experiments \cite{Budker2002RMP}.}. The stronger the light, the more pronounced is the effect, and larger deviation from the $\pi/2$ phase shift is observed. Similarly, when the ground-state relaxation rate $\Gamma_g$ is comparable to or higher than the magnetic-field modulation $\Delta\omega_L$, the time scales of the Larmor precession and relaxation become comparable, and again the phase-shift delay is reduced [the green line in Fig.~\ref{fig:ParametricLowPhaseAmpl}(b)]. 
Figure~\ref{fig:ParametricLowPhaseAmpl}(a) additionally shows that the highest medium anisotropy could be achieved for a small ground-state relaxation rate $\Gamma_g$. However, it also shows that for a larger $\Gamma_g$ the range of Rabi frequencies $\Omega$ leading to an effective medium-anisotropy generation is wider and the maximum signal amplitude occurs for higher $\Omega$ [green and red lines in Fig.~\ref{fig:ParametricLowPhaseAmpl}(a)].
\subsection{PNMOR with strong DC fields\label{sec:NMORstrong}}

\begin{figure}[ht]
\includegraphics[width=\columnwidth]{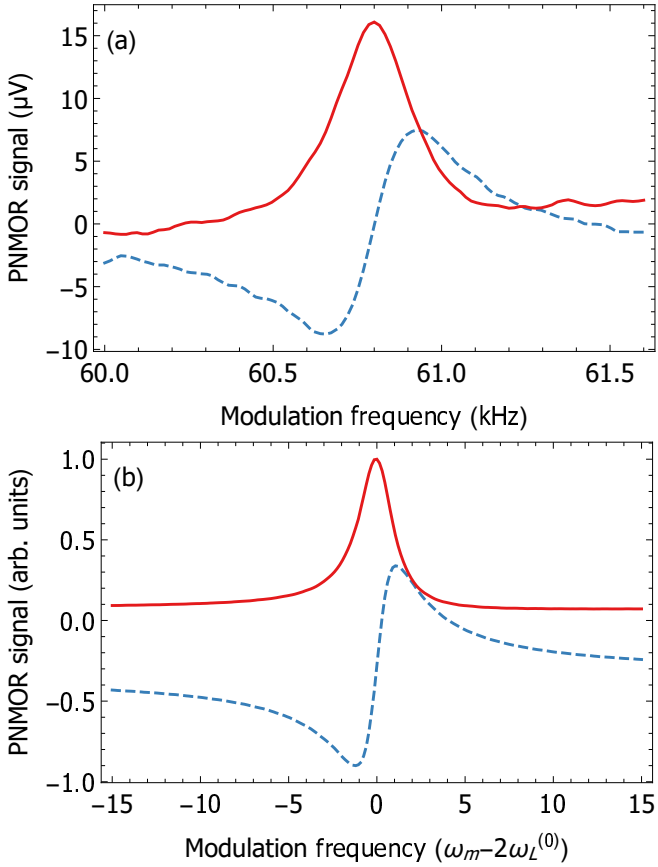}
\caption{(Color online) In-phase (dashed blue) and quadrature (solid red) component of (a) measured and (b) simulated PNMOR signal versus the modulation frequency $\omega_m$ (in plot (b), the modulation-frequency axis is shifted by $2\omega_L^{(0)}$). The experiment was performed for $\omega_L^{(0)}/2\pi\approx\Delta \omega_L/2\pi\approx 3\times 10^4$~Hz ($\sim 43$~mG) and laser power of $\sim20$~$\mu$W. The original experimental data was shifted by $+145$~$\mu V$ (in-phase component) and $-20$~$\mu V$ (quadrature component). For numerical calculations, following values of parameters were used: $\Gamma_e/\Gamma_g=10^{4}$, $\Omega/\Gamma_g= 1.06\times 10^{2}$ and $\Delta\omega_L/\Gamma_g=\omega_L^{(0)}/\Gamma_g=6\times 10^{2}$.}
\label{fig:ParametricHighExp}
\end{figure}

\begin{figure}
\includegraphics[ width=\columnwidth]{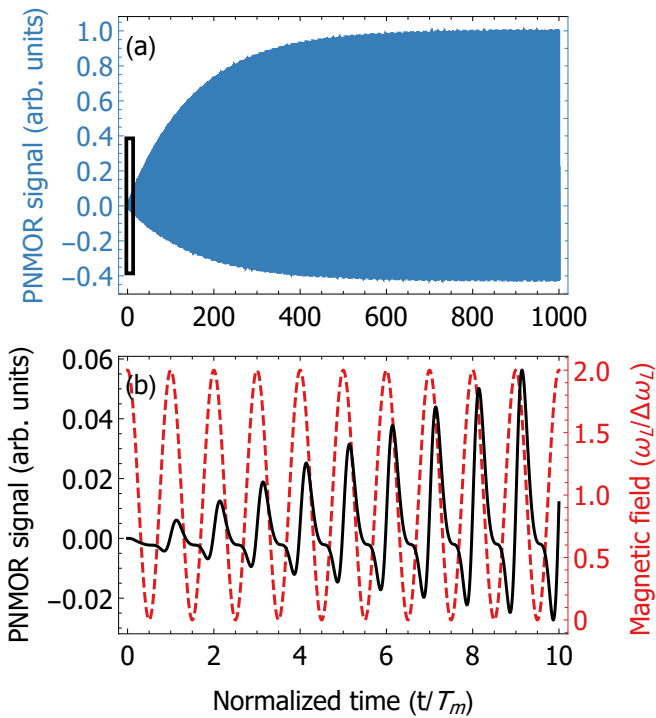}

\caption{(Color online)  Simulated dependence of PNMOR signal on time. (a) PNMOR signal over ${1000~T_{m}}$  demonstrating reaching the quasi-steady by the system. The black rectangle corresponds to evolution over ${10~T_{m}}$, presented in detail in plot (b). (b) PNMOR signal (solid black) and the modulated magnetic field (dashed blue) over ${10~T_{m}}$. The simulations were performed for parameters the same as in Fig.~\ref{fig:ParametricHighExp}.}
\label{fig:OpticalCoherencesHigh}
\end{figure}

The next step is the analysis of PNMOR in strong DC fields ($\omega_L^{(0)}\gg\Gamma_g$). The in-phase and quadrature components of the PNMOR signal recorded as a function of the magnetic-field modulation frequency $\omega_m$ for the comparable amplitudes of both field components ($\Delta\omega_L/2\pi\approx 3\times10^4$~Hz, $\omega_L^{(0)}/2\pi\approx 3\times10^4$~Hz) are plotted in Fig.~\ref{fig:ParametricHighExp}(a).
As shown, operation under such conditions enables recording of dispersive (in-phase) and absorptive (quadrature) resonances centered at $60.8$~kHz, corresponding to twice the Larmor frequency of the DC-field component $\omega_L^{(0)}$.
The original experimental data contained frequency-independent offsets \footnote{caused most likely by  a cross-talk between the field induced by the solenoid and the photodiode preamplifier mounted inside it}, but for better visibility both components were offset vertically. Figure~\ref{fig:ParametricHighExp}(b) presents results of numerical simulations, performed with parameters corresponding to those in experiment, which supports our explanation of the phenomenon. Similarly, as before, dynamic signals (such as shown in Fig.~\ref{fig:OpticalCoherencesHigh}) are integrated over a single modulation period within the quasi-steady state. The shapes of the in-phase and quadrature components of the simulated signal are consistent with the experimental data [Fig.~\ref{fig:ParametricHighExp}(a)] with a slight difference in the quadrature-to-in-phase component amplitude ratio. We relate the difference to simplifications of our model. Increased width (350 $\mu$G) of the registered resonance is caused by inhomogeneity of the applied magnetic field. Similarly to the weak DC field case (Sec.~\ref{sec:NMORweak}), the amplitude of the PNMOR signal does not show any significant dependence on the amplitude $\Delta\omega_L$ of the AC component of the magnetic field. In general, at higher magnetic fields, the resonance from Fig.~\ref{fig:ParametricHighExp} may be split due to the nonlinear Zeeman effect \cite{Pustelny2011Tayloring}, however, in this experiment the splitting is much smaller than the resonance width.

\begin{figure}
\includegraphics[width=\columnwidth]{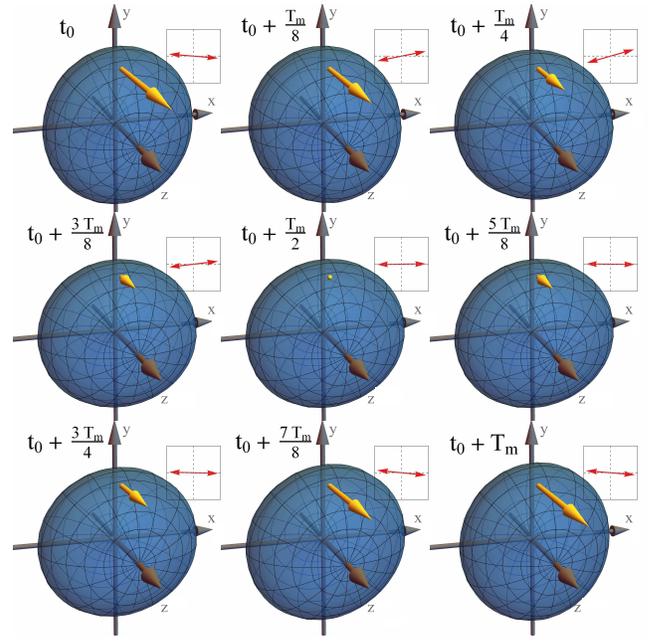}
\caption{(Color online) Full period of the quasi-steady-state evolution of the DM under interaction with (nearly) linearly polarized light and the magnetic field with large DC and AC components ($\Delta\omega_L=\omega_L^{(0)}$,  $\Delta\omega_L=6\times 10^{2} \Gamma_g$, $\Gamma_e=10^4 \Gamma_g$ and $\Omega=1.06 \times 10^{2} \Gamma_g$). The modulation frequency $\omega_m$ satisfies the resonance condition $\omega_m=2\omega_L^{(0)}$. The inset shows calculated polarization rotation (for better visibility, the rotation angle was increased 50-fold), while the yellow arrow represents the instantaneous magnetic-field vector.}
\label{fig:ParametricHigh1}
\end{figure}

Time evolution of the DM for the resonant case, ${\omega_m=2\omega_{L}^{(0)}}$,  upon a single cycle of the magnetic-field oscillation is presented in Fig.~\ref{fig:ParametricHigh1}. 
In contrast to the previously discussed case, here the anisotropy axis does not swing back and forth around the major axis of light polarization, but rotates in one direction determined by the DC-field component. This is due to the condition $\Delta\omega_L=\omega_L^{(0)}$ used for the simulations (for $\Delta\omega_L>\omega_L^{(0)}$ the precession reversal is observed). While in the standard NMOR, unidirectional precession and interaction with CW light washes out the transverse atomic polarization, in PNMOR the magnetic field modulation at $\omega_m=2\omega_L^{(0)}$ leads to appearance of the parametric resonance. The behavior may be explained and illustrated using the angular-momentum probability surfaces. Due to the magnetic field modulation, the ellipsoid spends more time with the orientation parallel to the light polarization (Fig.~\ref{fig:ParametricHigh1}, $t=t_0+0.5 T_m$) than in any other orientation (see, for example, Fig.~\ref{fig:ParametricHigh1}, $t=t_0$ and $t=t_0+T_m$). For parallel orientations, any freshly-pumped atoms contribute constructively to the preexisting atomic polarization and enhance the anisotropy, whereas in the case of other orientations they are smeared over all possibilities. Thereby, a specific orientation of the transverse alignment dominates over other orientations and the medium acquires a preferred polarization direction in the $xy$-plane. This phenomenon occurs only when the frequency of magnetic-field modulation $\omega_m$ is synchronized with the Larmor frequency of the DC-field component $\omega_L^{(0)}$ ($\omega_m=2 \omega_L^{(0)})$. When this condition is not fulfilled, polarization slows down with a different spatial orientation of the probability surface. This, over many modulation periods, leads to the averaging of the transverse polarization, similar to that observed in conventional NMOR (region III).

Figure~\ref{fig:OpticalCoherencesHigh} presents time-dependent polarization rotation simulated for the strongest DC component of the magnetic field.
Similarly as in Sec.~\ref{sec:NMORweak}, light polarization starts to rotate right after turning on the light and magnetic field modulation. The amplitude of these oscillations increases over time finally stabilizing at a parameter-dependent value. During its evolution, the precession periodically slows down and speeds up, spending, if the resonance condition ($\omega_m=2\omega_L^{(0)}$) is met, more time oriented along the major-axis of elliptically polarized light, which leads to the efficient generation of medium polarization. At the times of such slow-downs however, no instantaneous rotation is observed. When the Larmor precession speeds up, the non-zero rotation signal is observed (Fig.~\ref{fig:OpticalCoherencesHigh}). Since the evolution is non-harmonic, in addition to carrier frequency $\omega_m$, higher harmonics are also observed in the signal.
\begin{figure}
\includegraphics[ width=0.91 \columnwidth]{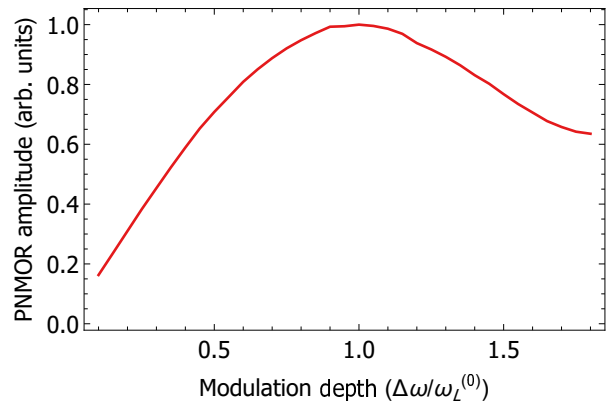}
\caption{Simulated dependence of the high field PNMOR signal amplitude on the depth of modulation. Apart from value of $\Delta \omega_l$, this simulation was performed for the same parameters as those used in Fig. \ref{fig:ParametricHighExp}.}
\label{fig:ModulationDepth}
\end{figure}
\\
Although experiment and simulations described thus far in this section discussed situation where the modulation depth of the strong DC field was almost exactly equal to the value of the DC field, it is worth pointing out that the signal amplitude of high-field PNMOR weakly depends on the magnetic field modulation depth, as presented in Fig. \ref{fig:ModulationDepth}. The optimal modulation depth fulfills condition $\Delta \omega \approx \omega_{L}^{(0)}$ and a 20\% deviation from the optimal modulation amplitude decreases the signal only by less than 10\%. This property of the high-field PNMOR resonance may be particularly useful for magnetic-field measurements in environments where background magnetic field is prone to drifts, i.e. for measurements in the Earth-field range.
\section{Conclusions \label{sec:Conlusions}}
In this paper we have demonstrated that the parametric resonance can be observed in NMOR experiments. In particular, we have identified two distinctive cases, where PNMOR can be observed; the weak and the strong DC magnetic fields regimes. We discussed the physical mechanisms responsible for creation of these resonances indicating the qualitative difference between these two regimes. The properties of simulated PNMOR resonances were found to be consistent with our measurements. Upon closer investigation of PNMOR resonance behaviour for a range of parameters we showed that the PNMOR with strong DC field is a good candidate for application in optical magnetometry, due to its wide dynamic range \footnote{This work was done in the magnetic fields reaching approx. 90 mG}, high sensitivity (resonance widths $<$ 50 $\mu$G), and partial immunity to background field drifts (as demonstrated by operation in semi-shielded environment). The magnetometric applications of PNMOR implicated in this paper will be presented elsewhere.
\begin{acknowledgments}
The authors would like to thank P. Anielski, D. Budker, M. Ledbetter and  B. Patton  for stimulating discussion. The work was supported from the grants of the National Science Centre, Poland within the OPUS programme (project no. 2015/19/B/ST2/02129). The contribution of PW is supported by the National Science Centre, Poland, project no. 2015/19/D/ST2/02195. The work of WG was supported by National Science Centre, Poland, project no. 2016/21/B/ST7/01430. PP would like to acknowledge the support of Polish Ministry of Science and Higher Education, project. no 7150/E-338/M/2018.
\end{acknowledgments}
\bibliography{sample}
\end{document}